\documentclass[twocolumn,useAMS,usenatbib]{mn2e}
\usepackage{natbib}
\usepackage{amssymb,amsmath}
\usepackage{graphicx,color}
\usepackage{hyperref}

\title[Sky obscuration as a weak lensing systematic]{Background sky obscuration by cluster galaxies as a source of systematic error for weak 
lensing}
\author[Simet \& Mandelbaum]{Melanie Simet$^1$\thanks{\tt msimet@andrew.cmu.edu}, 
Rachel Mandelbaum$^1$
\\$^1$McWilliams Center for Cosmology, Wean Hall, Carnegie Mellon University, 5000 Forbes Ave, Pittsburgh PA, USA, 15213}

\date{\today}

\begin{document}
\maketitle

\begin{abstract}
Lensing magnification and stacked shear measurements of galaxy clusters rely on measuring the 
density of background galaxies behind the clusters. 
The most common ways of measuring this quantity ignore the fact that
some fraction of the sky is obscured by the cluster galaxies
themselves, reducing the area in which background galaxies can be
observed.  We discuss the size of this effect in the Sloan Digital Sky
Survey (SDSS) and the Canada-France-Hawaii Telescope Lensing Survey
(CFHTLenS), finding a minimum 1 per cent effect at $0.1h^{-1}$Mpc from the
centers of clusters in SDSS; the effect is an order of magnitude
higher in CFHTLenS.  The resulting biases on cluster mass and
concentration measurements are of the same order as the size of the
obscuration effect, which is below the statistical errors for cluster
lensing in SDSS but likely exceeds them for CFHTLenS.  We also
forecast the impact of this systematic error on cluster mass and
magnification measurements in several upcoming surveys, and find that
it typically exceeds the statistical errors.  We conclude that future
surveys must account for this effect in stacked lensing and
magnification measurements in order to avoid being dominated by
systematic error.
\end{abstract}


\section{Introduction}

Studies of galaxy clusters are powerful probes of cosmology \citep{2006astro.ph..9591A,
2011ARAA..49..409A} and dark matter physics \citep{2005RvMP...77..207V,2012ARAA..50..353K}, 
especially in combination with weak gravitational lensing measurements
\citep{2013PhR...530...87W}.  However, accurate calibration of the
cluster masses is critical, and requires a careful account of all
sources of systematic error.  Many studies have 
explored the impact of well-known sources of systematic uncertainty such as line-of-sight structure 
\citep{2010ApJ...709..286M,2011MNRAS.412.2095H}, photometric redshift errors 
\citep{2010MNRAS.401.1399B,2012ApJS..201...32S}, triaxiality 
\citep{2007MNRAS.380..149C,2012MNRAS.425.2287H,2012JCAP...05..030S}, and cluster centroiding 
\citep{2010MNRAS.405.2078M,2012ApJ...757....2G}, as well as systematics general to weak lensing 
such as shape measurement bias \citep{2002AJ....123..583B,2010MNRAS.405.2044B,2012MNRAS.423.3163K}, 
detector effects \citep{2013MNRAS.429..661M}, and intrinsic alignments 
\citep{2004PhRvD..70f3526H,2012JCAP...05..041B,2013MNRAS.432.2433H}.  
We discuss here 
the impact of one particular systematic, the obscuration of part of the background sky by galaxies 
within the clusters themselves due to the (in)ability of software to separate nearby galaxies 
(deblending).\footnote{Background galaxies may also be partially obscured by cluster dust, altering their
colors and fluxes without impacting their detection levels, but we do not address this source of
error in this work.} 

As discussed further in section~\ref{Theory}, magnification measurements (directly) and mass
inference via stacked shear 
measurements (indirectly) depend on the observed number density of sources behind the galaxy 
clusters \citep{2001PhR...340..291B, 2009PhRvL.103e1301S, 2004AJ....127.2544S}.  If the observed 
area is decreased due to an unaccounted-for obscuration by cluster galaxies, the number density 
errors can propagate back to the cosmological measurements of interest.  Perniciously, the number 
density of galaxies in a cluster is a radius-dependent quantity which may mix into the 
radius-dependent magnification and lensing signals.  The possibility of this obscuration effect has 
been noted before in \citet[\S 1.2]{2001PhR...340..291B}, \citet[\S 3.5]{2011ApJ...735..118R},
and \citet[\S 4.2]{2014MNRAS.439...48A},
and has now been detected in the Dark Energy Survey \citep{2005astro.ph.10346T} in 
\citet{2014arXiv1405.4285M}.  The corrections required for this effect in the 
\citeauthor{2014arXiv1405.4285M} cluster sample are as large as a multiplicative factor of 3 in the 
inner regions of one cluster, and are non-negligible at $\leq 1 h^{-1}$Mpc for all clusters
studied.  We discuss this important measurement more in section \ref{Forecasts}.

We present a measurement of this effect in the Sloan Digital Sky 
Survey \citep[SDSS;][]{2000AJ....120.1579Y} and explore how it should
differ for the Canada-France-Hawaii Telescope Lensing Survey 
\citep[CFHTLenS;][]{2012MNRAS.427..146H}.  The CFHTLenS results are used to forecast
the size of this effect in upcoming deep surveys.  We discuss the theoretical underpinnings in section 
\ref{Theory}, describe the data sets in section \ref{Data}, show results in section \ref{Results}, 
and discuss implications for ongoing and future surveys such as the
Hyper Suprime-Cam Survey 
\citep{2012SPIE.8446E..0ZM}, 
the Dark Energy Survey, Euclid \citep{2011arXiv1110.3193L}, and 
the Large Synoptic Survey Telescope \citep[LSST;][]{2009arXiv0912.0201L} in section \ref{Forecasts}.

\section{Theory}\label{Theory}

Weak lensing is the distortion of background galaxy shapes due to matter in foreground lenses such 
as galaxies and clusters.  It is sensitive to the density profile of the lenses on small scales.  
A concentration of mass will distort the background galaxies to appear more tangentially 
aligned to the foreground lens than random galaxies would be, and so galaxy-galaxy or cluster-galaxy
lensing attempts to measure this preferential alignment.  In large surveys, where the signal to 
noise per foreground lens is small, lensing around galaxies or
clusters involves stacking: the shears from all 
pairs of lenses and background galaxies in a given radius bin for a
given lens sample are averaged 
to increase the signal-to-noise ratio.  With stacking, ongoing and future surveys are expected 
to measure shear profiles with percent-level errors or better.  Stacking also reduces the effect of 
some of the systematic errors, such as the large-scale structure contribution (which will have a 
different, uncorrelated sign and amplitude for the different lenses on
small scales) and centroiding (provided that the centroiding errors
can be characterized statistically for a large number of lenses).  

For most cosmological analyses, we are in the thin lens limit, where the lenses are far from the 
observer and also far from the objects being lensed.  In this case, we can describe the image 
distortions by a lensing potential,
\begin{equation}
 \psi(\boldsymbol{\theta}) = \frac{1}{\pi} \int_{\mathbb{R}^2} d^2 \theta' 
    \frac{\Sigma(\boldsymbol{\theta}')}{\Sigma_{\mathrm{cr}}} 
    \ln \left\vert \boldsymbol{\theta} - \boldsymbol{\theta}'\right\vert 
\end{equation}
which depends on the surface mass density $\Sigma(\boldsymbol{\theta}')$ and a term called the 
critical density, 
\begin{equation}
\Sigma_{\mathrm{cr}} = \frac{c^2}{4\pi G} \frac{D_s}{D_l D_{ls}},
\end{equation}
which encodes the strength of the lensing signal as a function of the (angular diameter) distances 
in the problem.  The ratio $ \Sigma(\boldsymbol{\theta}')/\Sigma_{\mathrm{cr}}$ is denoted $\kappa$ and 
called the convergence.  For axially symmetric lenses, the tangential
shear (i.e., the measured coherent shape distortion of background
galaxies) is
\begin{equation}
\gamma_{t}(R) = \bar{\kappa}(<R) - \kappa(R)
\end{equation}
and for non-axially symmetric lenses the same relation holds as long as shears and convergences are 
azimuthally averaged.  We work here in the weak lensing limit, where these induced shears are of 
order 10 per cent or less of the RMS shear due to the intrinsic galaxy
shapes. 

In addition to the shape distortions, the observed galaxies are made larger by a factor $1+\kappa$, 
which also increases the total observed flux (since lensing conserves surface brightness;  
\citealt{2001PhR...340..291B}).  The measured number density of background galaxies under magnification, relative to the 
number density of background galaxies in the field, is increased due to the greater detectability of brighter and larger 
objects but decreased due to the overall number density reduction caused by the dilation of the 
background sky.  If we assume the selection function for the galaxies included in the survey is a 
step function (size $r>r_{\mathrm{min}}$, apparent magnitude
$m<m_{\mathrm{min}}$ or flux $f<f_{\mathrm{min}}$), we can write the observed number density as 
\citep{2009PhRvL.103e1301S}
\begin{equation}
n_{\mathrm{obs}}(R) = n_0[1 + (2\beta_f + \beta_r - 2)\kappa(R)]
\end{equation}
with
\begin{equation}\label{magnification-parameters}
\begin{split}
\beta_r & = \left. -\frac{\partial \ln n_{\mathrm{obs}}}{\partial \ln r}
             \right\rvert_{\substack{m<m_{\mathrm{min}} \\ r=r_{\mathrm{min}}}} \\ 
\beta_f & = \left. -\frac{\partial \ln n_{\mathrm{obs}}}{\partial \ln f}
             \right\rvert_{\substack{r>r_{\mathrm{min}} \\ f=f_{\mathrm{min}}}} = 
             \left. 2.5\frac{\partial \ln n_{\mathrm{obs}}}{\partial m}
             \right\rvert_{\substack{r>r_{\mathrm{min}} \\ m=m_{\mathrm{min}}}}.
\end{split}
\end{equation}
In practice, the number density is weighted to account for measurement errors in magnitude and 
photo-$z$.  
The partial derivatives above should be taken relative to the weighted
(not raw) number density, which can make a significant difference in the measured parameter values.

Since a magnification measurement using number counts compares the
observed surface number density of objects 
 with some expected value for the field, 
the fraction of the sky that is obscured by cluster galaxies is a
direct source of error.  What one actually measures (when neglecting
obscuration) is
\begin{eqnarray}
\hat{n}_{\mathrm{obs}}(R) &=& (1-f_{\mathrm{obsc}})n_0(R)\left[1 +
  (2\beta_f + \beta_r - 2)\kappa(R)\right]\\
&=&(1-f_\text{obsc})n_\text{obs}(R).\label{eq:nobs}
\end{eqnarray}
Ideally, the convergence can be estimated via 
\begin{equation}
\kappa = \frac{n_\text{obs}-n_0}{n_0 q}\equiv \frac{\delta n}{q}
\end{equation}
where we define $q\equiv 2\beta_f + \beta_r-2$ and galaxy number
overdensity $\delta n \equiv n_\text{obs}/n_0-1$.  
When we use the incorrect estimator for $\hat{n}_\mathrm{obs}$ that
does not account for the obscuration effect, Eq.~\ref{eq:nobs}, the
estimated convergence becomes 
\begin{eqnarray}
\hat{\kappa}(R) &=& \frac{\hat{n}_{\mathrm{obs}}(R)/n_0(R)-1}{q} \\
&=&
\frac{(1-f_\text{obsc})n_\mathrm{obs}(R)/n_0(R)-1}{q}\label{magnification-equation}\\
&=& (1-f_\text{obsc})\kappa(R) - \frac{f_\text{obsc}}{q}
\end{eqnarray}
which is smaller than the true value.

For modeling $\kappa(R)$ we use a spherically symmetric Navarro-Frenk-White (NFW) density profile, 
which fits well with both simulations and previous measurements and which has an analytic solution 
\citep{2000ApJ...534...34W}.  We use a mass (defined using a spherical
overdensity of $200\rho_\text{crit}$) of $10^{14} h^{-1} M_{\odot}$ and a concentration $c_{200}=4$. 

In principle, the estimated lensing shears $\gamma_{1,2}$ are independent of the obscuration of the 
background galaxies; they are, after all, typically estimated using
averages of the observed galaxy shears.  In practice, 
however, a correction must be made due to the use of photometric redshifts (photo-$z$s), and this 
correction relies on the number density of objects \citep{2004AJ....127.2544S}.  The strength of 
the lensing signal depends on the redshifts through $\Sigma_{\mathrm{cr}}^{-1}$, which is 0 for 
$z_{\mathrm{source}} \leq z_{\mathrm{lens}}$ and positive for $z_{\mathrm{source}} > 
z_{\mathrm{lens}}$.  Photo-$z$s scatter physically associated (and therefore unlensed) galaxies in 
redshift space, so some unlensed galaxies with no expected signal will be included in the averaging
intended to find the lensing signal, reducing the amplitude of the measurement.  Near galaxy 
clusters--which contain many galaxies at the same redshift as the lens--the effect can be large, up 
to multiplicative factors of order a few \citep{2007ApJ...669...21P,2009ApJ...703.2217S}.  The 
exact magnitude of the effect depends on the photo-$z$ quality and the typical line-of-sight 
separation between the sources and the lenses: higher photo-$z$ errors
or biases, and a large overlap between lens- and source-redshift
distributions, will both increase the effect.  A 
correction is usually made based on the weighted number density of objects around lenses compared to 
that around similarly distributed random points \citep{2004AJ....127.2544S}:
\begin{equation}
C(R) = \frac{N_{\mathrm{rand}}}{N_{\mathrm{lens}}}
       \frac{\sum_{\mathrm{lens}} w_\mathrm{lens}}{\sum_{\mathrm{rand}} w_\mathrm{rand}}
\end{equation}
for lens-background galaxy pairs and random point-background galaxy pairs at radius R, which ties the measured 
shear to the measured number density.  (Optimal weighting for this calculation involves not only the
uncertainty on the shear, but also a weighting by the critical density
through $\Sigma^{-2}_{\mathrm{cr}}$.)  Some authors
\citep[e.g.][]{2011ApJ...735..118R} include this correction directly 
in their cluster lensing shear estimator, rather than 
writing it as a separate correction factor, but both formalisms are attempting to account for the 
dilution of the lensing signal by physically associated sources, and
are therefore subject to obscuration effects.  

Given 
a factor $f_{\mathrm{obsc}}$, the fraction of the background sky obscured by galaxies within the galaxy 
cluster, the estimated $\hat{C}(R)$ while ignoring obscuration is
actually 
\begin{equation}\label{BoostFactor-obsconly}
\hat{C}(R) = (1-f_\text{obsc})C(R).
\end{equation}
However, ignoring the effect of magnification on the number counts
will also lead to an error in the measured $\hat{C}$:
\begin{equation}\label{BoostFactor}
\hat{C}(R) = (1-f_\text{obsc}) \left[1+q\kappa(R)\right]C(R).
\end{equation}
A slight subtlety is that the weights used to calculate the shear
signals often differ from those used in estimates of magnification, so
the relevant $q$ in this equation may differ from what should be used
in Eq.~\ref{magnification-equation}.  Since some studies have
already accounted for the effects of magnification on the number counts
\citep[e.g.,][]{2006MNRAS.372..758M}, we will directly compare the systematic error in
the boost factor and therefore shear signal using
Eq.~\ref{BoostFactor-obsconly} (obscuration effects alone)
vs. Eq.~\ref{BoostFactor} (obscuration effects and magnification),
using the appropriately-weighted $q$ for shear estimation.

\section{Data}\label{Data}

We measure the obscuration effect with two surveys: the Sloan Digital Sky Survey 
\citep[SDSS;][]{2000AJ....120.1579Y} and the Canada-France-Hawaii Telescope Lensing Survey 
\citep[CFHTLenS;][]{2012MNRAS.427..146H}.  SDSS is wide but comparatively shallower, which will 
give us an idea of the problem in current stacked weak lensing measurements; CFHTLenS has a small area but 
is deeper, which will allow us to project the magnitude of the effect for upcoming surveys (section
\ref{Forecasts}).  Since magnification measurements are usually made on photometric samples, and shear
measurements are usually made on shape samples that include additional
selection criteria and/or weight factors, we will show results from both types of catalogue.

\subsection{SDSS source catalogues}

\begin{figure}
  \begin{center}
  \includegraphics[width=0.45\textwidth]{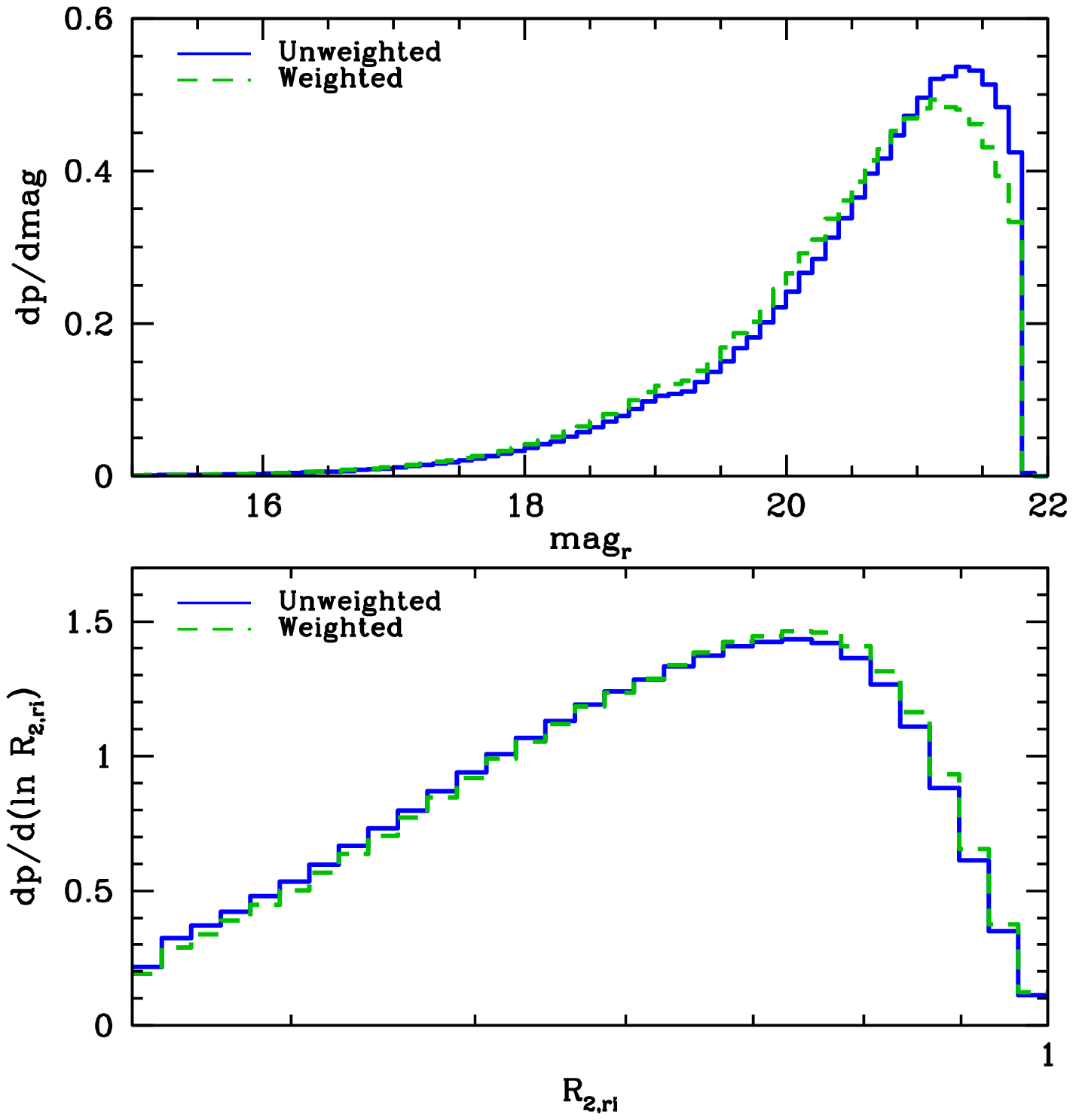} 
  \caption{Weighted and unweighted distributions
    $\mathrm{d}p/\mathrm{d}\mathrm{mag}$ and $\mathrm{d}p/\mathrm{d}
    \log R_{2,ri}$ measurements for the SDSS shape catalogue, necessary for 
            magnification calculations (Eq. \ref{magnification-equation}).}\label{number-sdss}
  \end{center}
\end{figure}

The shape catalogue from SDSS is described in \citet{2012MNRAS.425.2610R}.  It contains 39 million 
galaxies in an area of 9243 deg$^2$ which pass photometry, shape noise and photo-$z$ cuts, for an 
average density of 1.2 galaxies/arcmin$^2$.  The limiting magnitude is $r<21.8$.  Typical seeing is 
$1.2\pm 0.2$ arcsec.  The data is processed through the \verb/Photo/ v5\_6 pipeline \citep{2001ASPC..238..269L,2002AJ....123..485S,2011ApJS..193...29A}.  Star-galaxy separation is performed
using the 
\verb/OBJC_TYPE/ flag, which compares extended galaxy model fit magnitudes with PSF magnitudes.  
Shapes are then measured with the re-Gaussianization PSF correction software 
\citep{2003MNRAS.343..459H}. We do not use the shape measurements directly here, but we 
do use the errors on the shape 
measurement to define galaxy weights for the boost factor calculation (Eqn. \ref{BoostFactor}).  The
weights also include the per-component intrinsic dispersion of galaxy shapes:
\begin{equation}\label{WeightEquation}
w_i = \frac{1}{e_{\mathrm{rms}}^2 + \sigma_{e,i}^2}.
\end{equation} 
where, for this catalogue, we take $e_{\mathrm{rms}} = 0.365$, consistent with the findings of 
\citet{2012MNRAS.425.2610R}.  In addition, the traces of the moment matrices from the adaptive 
moments for the PSF ($P$) and image ($I$) are used to define a resolution factor,
\begin{equation}\label{resolution-equation}
R_2 = 1 - \frac{T_P}{T_I}
\end{equation}
We use this as the size parameter for our magnification calculation with the SDSS shape (not photometric) catalogue, 
rather 
than, e.g., a measured galaxy radius, because this is the quantity
that was used for galaxy selection.
As discussed later in Section \ref{Results}, we will also use it to determine an average PSF-convolved 
galaxy size.  
In practice, the cutoff for good shape measurement (and thus 
inclusion in our shape catalogue) is that $R_2>1/3$ in both $r-$ and $i-$band imaging, so 
functionally we use \citep{2005MNRAS.361.1287M}
\begin{equation}
R_{2,ri} = \min(R_{2,i},R_{2,r})
\end{equation}
as the size parameter for our analysis.  The weighted distributions $\mathrm{d}N/\mathrm{d}\mathrm{mag}$ and 
$\mathrm{d}N/\mathrm{d} \log R_{2,ri}$, 
necessary for magnification, are shown in Fig. \ref{number-sdss}, and the derived parameters are 
given in Table~\ref{magnification-table}.  More details on the adaptive moments 
and the resolution calculation can be found in
\citet{2005MNRAS.361.1287M} and \citet{2012MNRAS.425.2610R}.  

\begin{table}
\begin{center}
\begin{tabular}{l|ll|c|}
\hline
 & \multicolumn{2}{|c|}{Shape} & Photometric \\
 & \multicolumn{1}{c}{$\boldsymbol{\beta}_f$} & \multicolumn{1}{c}{$\boldsymbol{\beta}_r$} & $\boldsymbol{\beta}_f$ \\
\hline
 SDSS unweighted & 1.1 & 0.22 & 1.9 \\
 SDSS weighted & 0.89 & 0.19 & 3.1 \\
 CFHTLenS unweighted & 1.3 & 0.14 & 1.8 \\
 CFHTLenS weighted & 0.71 & 0.049 & 2.1\\
\hline
\end{tabular}
\end{center}
\caption{Parameters for the magnification calculation, Eq. \ref{magnification-equation}, as defined
         in Eq. \ref{magnification-parameters} and calculated using the measurements in Fig. 
         \ref{dndzfig}.  These correspond to the slopes of log(galaxy counts) relative to 
         the magnitude ($\boldsymbol{\beta}_f$) and galaxy size ($\boldsymbol{\beta}_r$), up to
         multiplicative factors described in Eq. \ref{magnification-parameters}.  
         Magnitude parameters were evaluated for the bin centred at $\mathrm{mag}_{r}=21.725$
         (SDSS) and $\mathrm{mag}_{i}=24.575$ (CFHTLenS), as sharp drops in $dN/d\mathrm{mag}$
         were observed beyond 
         those points, although a formal hard cutoff at the upper edges of those bins was not 
         imposed in the rest of the analysis.  The size parameters were both evaluated in the 
         smallest bin included in the catalogue.  Since there is no size cutoff in the photometric
         samples, we do not consider size magnification in that
         case.}\label{magnification-table}
\end{table}

For the corresponding photometric catalogue, we use a superset of the shape catalogue.  We include 
the galaxies that did not pass the resolution cut (and have no measured shapes) in addition to the
galaxies of the shape catalogue.  For weights, we use the $r$-band photometric error plus a small intrinsic
dispersion:
\begin{equation}\label{MagWeightEquation}
w_i = \frac{1}{0.02^2 + \sigma_{\textrm{mag},i}^2}.
\end{equation} 

\begin{figure}
  \begin{center}
  \includegraphics[width=0.45\textwidth]{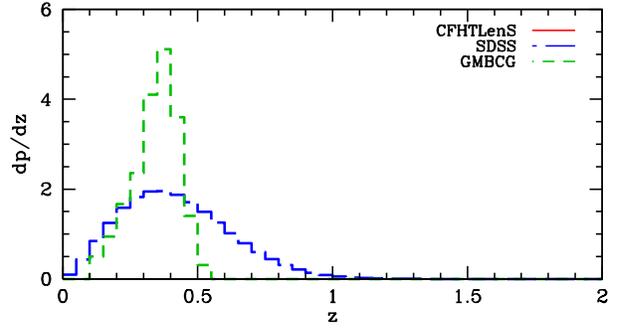} 
  \caption{Unweighted redshift distributions ($\mathrm{d}p/\mathrm{d}z$) for the SDSS and CFHTLenS shape catalogues and GMBCG cluster 
           catalogue. Photometric catalogues have a similar redshift distribution, with a slightly more populated high-redshift tail and a slightly less populated peak, but all such changes are at higher redshifts than the cluster sample.}\label{dndzfig}
  \end{center}
\end{figure}

The photometric redshifts for this catalogue were generated with the Zurich Extragalactic Bayesian 
Redshift Analyzer \citep[ZEBRA;][]{2006MNRAS.372..565F}.  Using the parameters described in 
\citet{2012MNRAS.420.3240N}, ZEBRA makes a point estimate of the photo-$z$ by fitting a set of 
templates to the observed galaxy colors (defined using corrected magnitudes from the PHOTO 
pipeline); the templates used here are from \citep{2000ApJ...536..571B} and consist of four observed 
galaxy SEDs, two synthetic blue spectra, and 5 interpolated spectra in between each pair of the 6 
spectra, for a total of 31 templates.  
The redshifts were validated against a set of spectroscopic and high-quality photometric redshifts
with the same selection function as the source sample, 
which resulted in rejecting starburst-type galaxies (10 per cent of the sample) due to unreliability.  The 
remaining galaxies have a small bias at $z \gtrsim 0.4$ which can be corrected for during lensing 
calculations through knowledge of the true $\mathrm{d}N/\mathrm{d}z$.  
The main effect in this analysis is the introduction of scatter in the optimal galaxy weighting 
(which depends on $\Sigma_{\mathrm{cr}}$), which will reduce our signal-to-noise ratio.  Further 
details of 
the photo-$z$ catalogue can be found in \citet{2012MNRAS.420.3240N}.  The best-fit parameterized 
$\mathrm{d}N/\mathrm{d}z$ from that paper for 
the SDSS shape catalogue can be found in 
Fig. \ref{dndzfig}; the functional form for the SDSS shape catalogue is a good fit for both the 
shape and photometric samples, although with slightly different
parameter values.

\subsection{SDSS cluster catalogue}

Our cluster catalogue is the GMBCG catalogue \citep{2010ApJS..191..254H}.  This catalogue 
contains approximately 51,000 galaxy clusters within the area of our SDSS shape catalogue.  Clusters
are chosen by fitting 
the color distribution of galaxies in a given patch of sky with a mixture of Gaussians; the field 
has a single broad Gaussian in color space, while clusters have an additional narrower Gaussian 
corresponding to the cluster red sequence.  The red-sequence galaxies are then convolved by a 
projected NFW kernel to define a density, used as an estimator for the clustering strength at that 
location.  Candidate brightest cluster galaxies \citep[BCGs, found at the centers of many clusters; 
see, e.g.,][]{2012ARAA..50..353K} are ranked by clustering strength, and then ``percolated'' down by 
removing candidate BCGs that appear in the clusters of more highly ranked BCG candidates.  The 
cluster richness is then estimated in a two-step process of measuring cluster galaxies in a fixed 
aperture, then using that number to define a more appropriate radius and measuring the number of 
galaxies within the new aperture.  The redshift distribution of these clusters is shown in
Fig. \ref{dndzfig}.  We generate random points with the same area
coverage as this catalogue using the SDSSPix  
software\footnote{\url{http://dls.physics.ucdavis.edu/~scranton/SDSSPix/}}.

\subsection{CFHTLenS source catalogues}

\begin{figure}
  \begin{center}
  \includegraphics[width=0.45\textwidth]{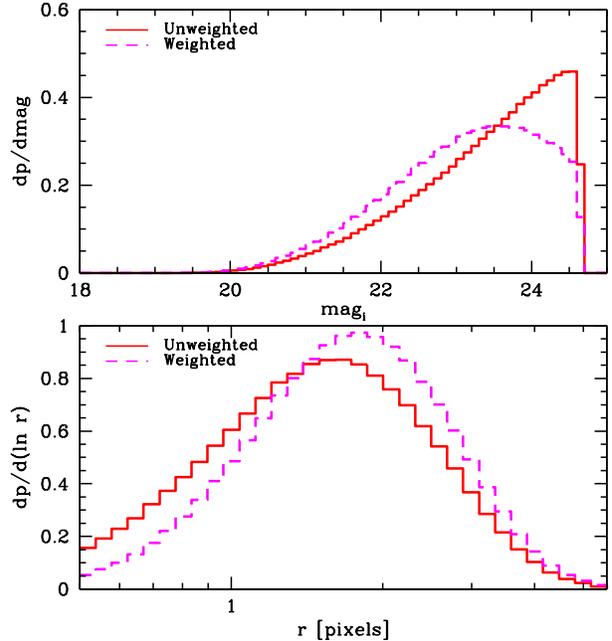} 
  \caption{Weighted and unweighted distributions
    $\mathrm{d}p/d\mathrm{mag}$ and $\mathrm{d}p/d \log r$
    measurements for the CFHTLenS catalogue, necessary 
            for magnification calculations (Eq. \ref{magnification-equation}).}\label{number-cfhtlens}
  \end{center}
\end{figure}

We use the public data catalogue from the CFHTLenS project\footnote{\url{http://cfhtlens.org/}}
\citep{2012MNRAS.427..146H}. The 
catalogue consists of 22.7 million 
objects in the four fields of the CFHT Legacy Survey (CFHTLS), covering a total of 154 square 
degrees with a limiting magnitude of 24.7 and average seeing of $(0.6-0.65)\pm 0.1$ arcsec depending 
on field, both in the stacked $i'$-band.  The stacked images contain fainter objects, but the 24.7 
mag cut is imposed due to noise bias considerations for shape measurement and a lack of fainter 
photo-$z$ training objects.  The data-analysis pipeline, THELI, is described in 
\citet{2013MNRAS.433.2545E}.  In brief, objects are detected by running SExtractor 
\citep{1996AAS..117..393B} on a full-depth $i'$-band stack, and then shears are measured using the 
\textit{lens}fit pipeline \citep{2013MNRAS.429.2858M}, which fits the object shapes simultaneously and 
jointly in the individual exposures using a Bayesian forward-modelling algorithm.  The models used 
are two-component (bulge and disk) galaxy models, with the size and ellipticity as free parameters, 
marginalized over centroid position.  The pipeline also generates a weight that includes the 
intrinsic shape dispersion as well as the \textit{lens}fit measurement error, which we can use directly
rather than making a calculation such as Eq. \ref{WeightEquation} for the measurements on the shape
catalogue.  For star-galaxy separation, we use the SExtractor \verb/CLASS_STAR/ for the photometric 
catalogue (where \textit{lens}fit may not have run successfully) and the \textit{lens}fit \verb/fitclass/ for the 
shape catalogue. 
Finally, for each
object, a scalelength along the major axis is produced, with a hard lower cutoff at $r=0.5$ pixels. 
This is the quantity we will use for the size magnification parameter in the shape catalogue.  
After cutting to 
objects classified as galaxies, we obtain a photometric catalogue of 15 million galaxies; after 
further removing galaxies with a \textit{lens}fit weight of 0 and using the \textit{lens}fit star-galaxy separation, 
7.5 million galaxies remain (an unweighted number density of 13.5 galaxies/arcmin$^2$) in our shape
catalogue. The measurements of 
$\mathrm{d}N/\mathrm{d}\mathrm{mag}$ and 
$\mathrm{d}N/\mathrm{d} \log r_{\textrm{FWHM}}$, necessary for magnification, are shown in Fig. \ref{number-cfhtlens}, and the 
derived parameters are given in Table~\ref{magnification-table}.  The redshift distribution is 
shown in Fig.~\ref{dndzfig}.

CFHTLenS photo-$z$s are measured using the Bayesian Photometric Redshift Code 
\citep[BPZ;][]{2000ApJ...536..571B,2006AJ....132..926C} with colors measured using SExtractor in 
dual image mode 
on homogeneous Gaussianized versions of the stacked images in all bands 
\citep{2012MNRAS.421.2355H}.  
BPZ estimates a posterior likelihood by comparison with a set of templates; here the templates used 
are from \citet{2004PhDT.........4C}.  
Comparison with spectroscopic and high-quality photometric redshifts shows strong agreement between 
the summed BPZ $P(z)$s and the true redshift distribution, even for fairly narrow bands in peak BPZ 
redshift, for $0.2<z_{\mathrm{BPZ}}<1.3$; redshifts beyond that range are significantly 
more uncertain.  We 
use this redshift range in the rest of the analysis, as it is low redshift enough to detect 
galaxies 
in our cluster lensing sample, but also deep enough to investigate the magnification signal behind 
the clusters.

\subsection{Cluster galaxy counts}\label{galaxy-counts}

Due to the conservative deblending settings used in the CFHTLenS pipeline, we cannot measure the obscuration 
effect well at $\lesssim 0.5 h^{-1}$Mpc, as shown in section~\ref{Results}.  Instead, we will
use a scaled version of the measurement from SDSS, using a cluster luminosity function and our 
knowledge of the depths of both surveys.  

\begin{figure}
  \begin{center}
  \includegraphics[width=0.45\textwidth]{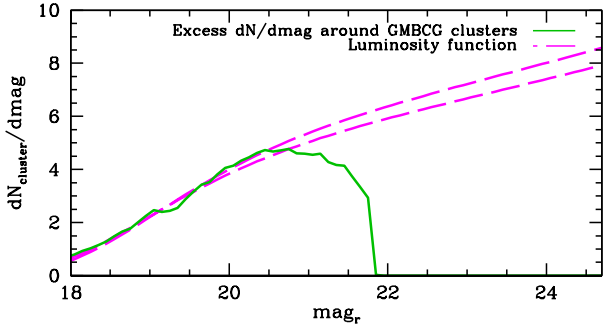} 
  \caption{Redshift distribution $\mathrm{d}N/\mathrm{d}\mathrm{mag}$
    measurements for galaxies in SDSS located in areas near GMBCG
    galaxy clusters, after the field galaxy 
            $\mathrm{d}N/\mathrm{d}\mathrm{mag}$ has been subtracted.  Two sample luminosity function fits are shown,
            which differ in amplitude but are similar to $\sim 10$ per
            cent under integration to the CFHTLenS limiting
            magnitude of 24.7.}\label{cluster-luminosity}
  \end{center}
\end{figure}

Shown in Fig. \ref{cluster-luminosity} is the $\mathrm{d}N/\mathrm{d}\mathrm{mag}$ distribution of SDSS shape catalog galaxies in 
an $r_{200}$ annulus around the GMBCG clusters with the average $\mathrm{d}N/\mathrm{d}\mathrm{mag}$ from the 
whole catalogue subtracted (after scaling to the appropriate 
area).  It is obviously different from the field distribution, which can be seen by comparing with 
Fig. \ref{number-sdss}.  For convenience, we fit a Schechter luminosity function 
\citep{1976ApJ...203..297S}
of the form
\begin{equation}\label{clf}
\frac{\mathrm{d}N}{\mathrm{d}m} = 0.4 \ln(10) \phi^* 10^{0.4 (m^*-m)(\alpha+1)}\exp{[-10^{0.4(m^*-m)}]}
\end{equation}
to the apparent magnitudes using a least-squares fit, fixing $\alpha$ to -1.1 in agreement with
previous studies of the cluster luminosity function 
\citep[see, e.g. the list in][]{2011ApJ...734....3Z}.  This is not a true luminosity function, 
since it is a funtion of apparent magnitudes rather than absolute magnitudes or luminosities;
we use it here merely as a convenience, but these results should not be used to predict the 
luminosity function at other redshift ranges.

There is large covariance between the fitted $\phi^*$ and $m^*$, but altering the parameters in
tandem along the degeneracy direction results in similar predictions at fainter magnitudes.  Since the quantity
of 
interest is the extrapolated curve, not the exact details of the cluster luminosity function, 
choosing the best-fit measurement is sufficient for our purposes.  The selection of a
range of points to be fit by the luminosity function matters more--the low magnitude end is
dominated by the BCGs, which are not well fit by the luminosity function 
\citep{2011ApJ...734....3Z}, and the high
end turnover is dominated by our source selection and not the clusters themselves.  Still, varying
these endpoints within reasonable parameters gives similar results: extending the luminosity function
to the CFHTLenS magnitude limit of 24.7 gives $\approx 3$ times as many cluster galaxies as would be seen in SDSS.  
(A sample high and low fit are shown in Fig. \ref{cluster-luminosity}.)  We will therefore use the 
excess number density from the SDSS catalogue, times three, for our cluster galaxy distribution in 
CFHTLenS, though we will present some additional tests that validate this procedure as
well.  We assume that the cluster galaxies at fainter magnitudes have the same spatial 
distribution as the bright galaxies, which is in line with previous studies finding no luminosity 
segregation in clusters \citep[e.g.][]{2005MNRAS.364.1147P}. 

We will also show, for comparison, some measurements around GMBCG clusters in the CFHTLenS region.  
These were masked using a simple geometric mask fit to the CFHTLenS area, which is 
made of sums of rectangles.  We use the Mangle software \citep{2008MNRAS.387.1391S} to generate randoms 
in the overlap regions between the GMBCG mask (represented as simple polygons, a small approximation 
given the small spatial extent of the CFHTLenS mask) and the CFHTLenS mask.  

\section{Results}\label{Results}

First, we must measure $f_{\mathrm{obsc}}$, the fraction of sky obscured by the galaxies within 
the cluster.  
This is the product of the number density of galaxies within the cluster $n_{cl}$ and 
the average obscured area per galaxy $A_{\mathrm{gal}}$.  The number density of galaxies within 
the cluster is simple: 
we measure the galaxy density around random points and around clusters; the observed number density 
is the number density of cluster galaxies minus the number density of non-cluster galaxies which 
can be detected in the same region.  This is the number density around random galaxies 
suppressed by the obscuration of the cluster and by the magnification.  
\begin{equation}\label{fobsc-equation}
\begin{split}
n_{\mathrm{cl}} & = n_{\mathrm{obs}} - (1+q\kappa) n_{\mathrm{rand}}(1-n_{\mathrm{cl}}A_{\mathrm{gal}}) \\
n_{\mathrm{cl}} & = \frac{n_{\mathrm{obs}} - (1+q\kappa) n_{\mathrm{rand}}}{1-(1+q\kappa)n_{\mathrm{rand}}A_{\mathrm{gal}}}
\end{split}
\end{equation}
We use $n_{\mathrm{rand}}$ and not $n_{\mathrm{bg}}$ since 
blending effects can remove both foreground and background galaxies, and
since the magnification factor $q\kappa$ is 0 for foreground galaxies (which must be accounted
for when measuring an average $\kappa$ or $\Sigma_{\mathrm{cr}}^{-1}$ for all galaxies in the 
sample). 
In a real analysis, this will need to be solved iteratively or jointly for $\kappa$ and the cluster 
mass.  However, as we will show shortly, $1+q\kappa$ is nearly 1 in
SDSS for the scales on which we are making this measurement, so we will not include it in
our measurement here.  

\begin{figure}
  \begin{center}
  \includegraphics[width=0.45\textwidth]{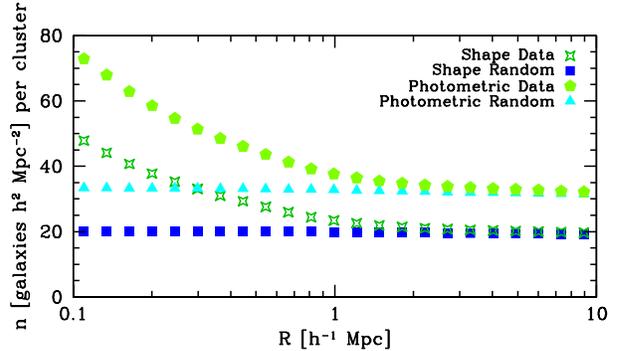} 
  \caption{Number density of SDSS galaxies at all redshifts around GMBCG clusters and random points.
           Both the shape and photometric catalogues are shown.  As expected, the number density is
           largest at the smallest radii.  Error bars are smaller than the point size at all radii
           and are omitted for clarity.}\label{ngals-sdss}
  \end{center}
\end{figure}

\begin{figure}
  \begin{center}
  \includegraphics[width=0.45\textwidth]{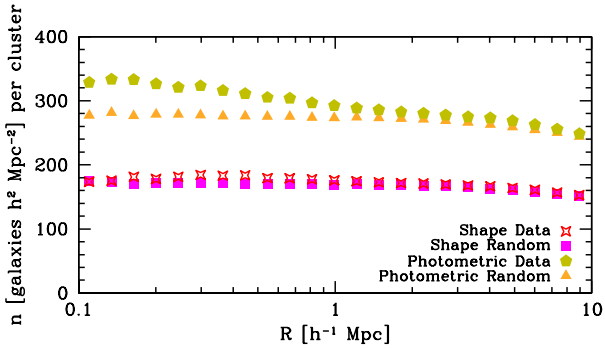} 
  \caption{Number density of CFHTLenS galaxies at all redshifts around GMBCG clusters and random 
           points.  Both the shape and photometric catalogues are shown.  In the photometric 
           catalogue, the number density increases towards small radii, although the magnitude
           of the change is smaller than expected.  In the shape catalogue, the effects of the
           conservative deblender can be seen as the turnover in the excess galaxy density around
           clusters, returning to the field level or even slightly
           below it for the smallest radii.}\label{ngals-cfhtlens}
  \end{center}
\end{figure}

\begin{figure}
  \begin{center}
  \includegraphics[width=0.45\textwidth]{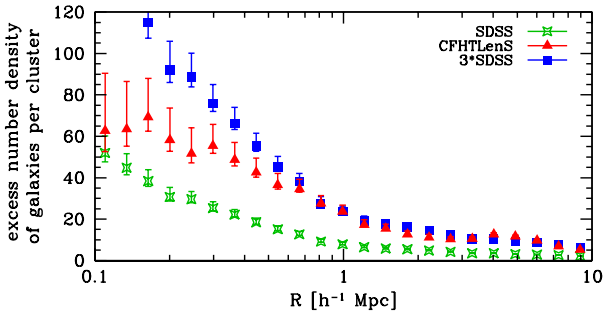} 
  \caption{The excess galaxy density around GMBCG clusters, as computed by subtracting the randoms 
            curve from the cluster curve in Figs.~\ref{ngals-sdss}~and~\ref{ngals-cfhtlens}.  
            Results from the photometric catalogues of SDSS and CFHTLenS are shown, as well as SDSS 
            scaled to the expected number of cluster member galaxies in CFHTLenS as described in 
            section~\ref{galaxy-counts}. We have used here a measurement of SDSS only around the
            427 clusters in the CFHTLenS region to remove any differences caused by the redshift and 
            richness distribution of the small CFHTLenS subsample of the GMBCG catalogue.  The 
            prediction based on SDSS holds up well at large radii, but the suppression of galaxy
            counts due to the data reduction can be seen at small
            radii.}\label{compare-ngals}
  \end{center}
\end{figure}

We show the measurements of $n_{\mathrm{obs}}$ and
$n_{\mathrm{rand}}$ and their difference in 
the shape and photometric catalogues for SDSS in Fig.~\ref{ngals-sdss} and in CFHTLenS in 
Fig.~\ref{ngals-cfhtlens}.  We also show a comparison of the difference between the two curves in
Fig.~\ref{compare-ngals}.  The suppression of galaxy density due to
the data reduction procedure in CFHTLenS can be seen
even at moderate radii, which cannot be explained by magnification (which should increase the 
galaxy density given the value of $q$).  Due to this difficulty, and to the increased stochasticity from the 
smaller number of clusters in the smaller CFHTLenS area, we will use a
scaled version of the curve from SDSS based on the cluster
luminosity function (as described in section~\ref{galaxy-counts}) for
the rest of our predictions involving the 
CFHTLenS shape catalogue.  The results in this figure above
$0.8h^{-1}$Mpc suggest that this rescaling factor is indeed valid.

To convert from the number of galaxies to the obscured fraction of the sky, we must characterize the 
fraction of sky obscured per cluster galaxy.  This obscuration comes
from the observed size of the galaxy after PSF convolution, and from
decisions made during the course of the deblending procedure.  As a
result, it varies 
from telescope to telescope and from data reduction method to data reduction method even when the 
galaxy population is identical.  We consider several possible choices
for the deblending-relevant area, covering a range that should capture
the characteristics of most pipelines.

For CFHTLenS, the 
deblending issue is the dominant concern for galaxy size. CFHTLenS shapes are measured on postage 
stamps of size $48\times 48$  
pixels (about $9\times9$\arcsec). No deblending attempts are made if a nearby galaxy is 
close enough 
that masking it out disturbs the isophotes of the target galaxy. Instead, the galaxy is excluded 
from the final catalogue. This procedure removes about 20 per cent of the galaxy sample.  The CFHTLenS team estimates 
this will bias lensing signals below 5\arcsec\ \citep{2013MNRAS.429.2858M}, or approximately half the 
width of the postage stamp, so we can assume the deblending-relevant galaxy size is something of
this order.  In comoving coordinates at the redshifts of the GMBCG
clusters, a length of 4.5\arcsec\  
corresponds to $20 h^{-1}$~kpc.  As an alternate statistic, we can also measure the apparent size of 
the PSF-convolved galaxies in this sample.  This is a redshift-dependent quantity, since the size of
the PSF will increase with angular diameter distance in either physical or comoving coordinates.
The full 
width at half max (FWHM) of the CFHTLenS galaxies, assuming a Gaussian core, is given in the shape 
catalogue, so we can measure the average comoving area of the galaxy+PSF within the FWHM circle.  We 
find the average area to be $36 h^{-2}\mathrm{kpc}^2$ for all CFHTLenS galaxies 
from 0.1-0.55 weighted (by BPZ maximum likelihood redshift) to match the redshift distribution of 
the GMBCG clusters, which corresponds to a radius of $3.4 h^{-1}$~kpc.  This is quite a 
bit larger than the average galaxy radius in the same redshift range, due to a combination of 
averaging $r^2$ rather than $r$ and the contribution of the PSF size.  

\begin{figure}
  \begin{center}
  \includegraphics[width=0.45\textwidth]{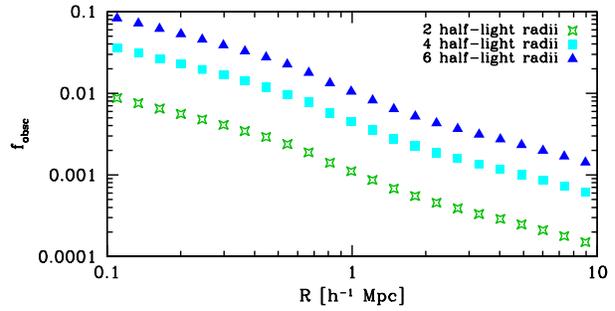} 
  \caption{Fraction of the sky obscured by cluster galaxies as a function of radius in SDSS for 
            several 
            choices of sky area per galaxy.  The curves are almost self-similar, but there are small 
            effects due to the correction of
            Eq.~\ref{fobsc-equation}.}\label{fobsc-sdss}
  \end{center}
\end{figure}

For SDSS, where the direct measurement of the combined galaxy$+$PSF FWHM was not saved post-processing, we use 
the resolution factor, Eq. \ref{resolution-equation}.  The traces of the moments matrices $T_I$ and 
$T_P$ are second-moment measurements, and should be roughly proportional to the area.  We can then 
use the measurement of the PSF FWHM at each galaxy position to write
\begin{equation}
A_{\mathrm{FWHM}} = \frac{\pi(\mathrm{FWHM}/2)^2}{1-R_{2}},
\end{equation}
where we use only the $r$-band measurements for simplicity.  
We find this area is $78 \times h^{-2}\mathrm{kpc}^2$, again matched to the
GMBCG redshift range, corresponding to a PSF-convolved half-light radius of $5.0 h^{-1}$~kpc.

\begin{figure*}
  \begin{center}
  \includegraphics[width=0.9\textwidth]{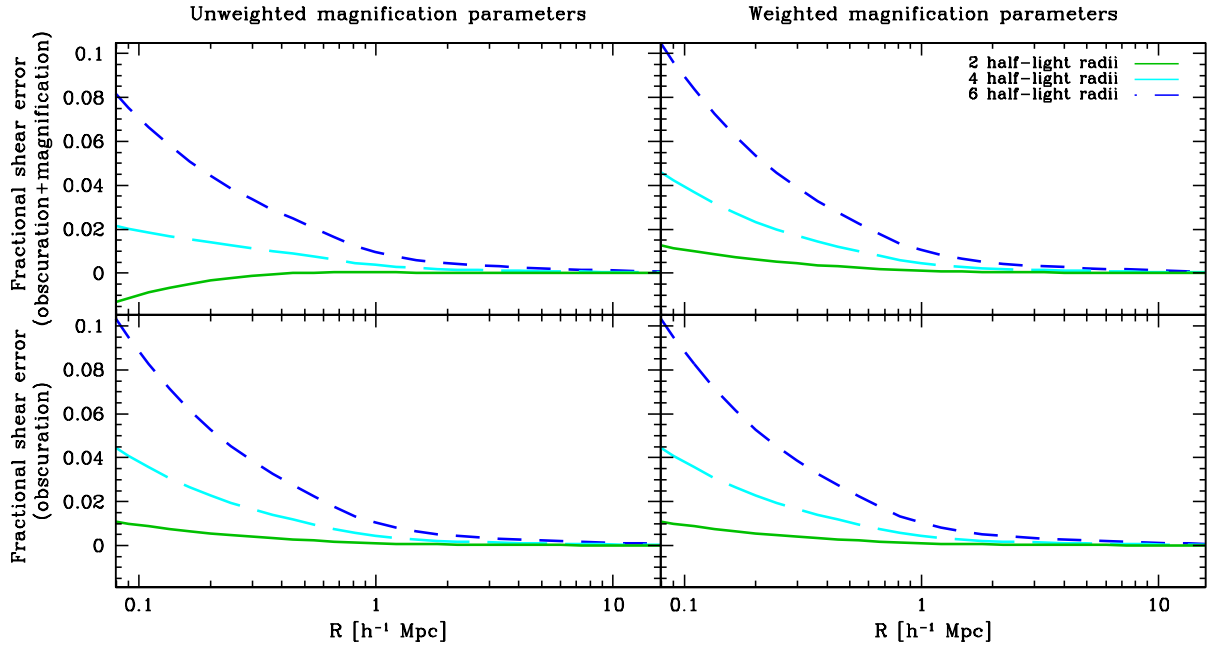} 
  \caption{Fractional errors on the shear in an SDSS-like survey due to neglecting magnification 
            and obscuration (top) or neglecting obscuration only (bottom).
            Magnification parameters derived using unweighted galaxy densities are shown 
            in the left-hand column, and using weighted densities in
            the right-hand column. 
            Galaxies magnified just above detection limits are so heavily downweighted in this 
            sample that the magnification has only a negligible effect when weights are included.  
            The obscuration can be quite large, up to order 10 per cent in the centers of the clusters 
            for conservative deblenders.}\label{shear-errors-sdss}
  \end{center}
\end{figure*}

\begin{figure*}
  \begin{center}
  \includegraphics[width=0.9\textwidth]{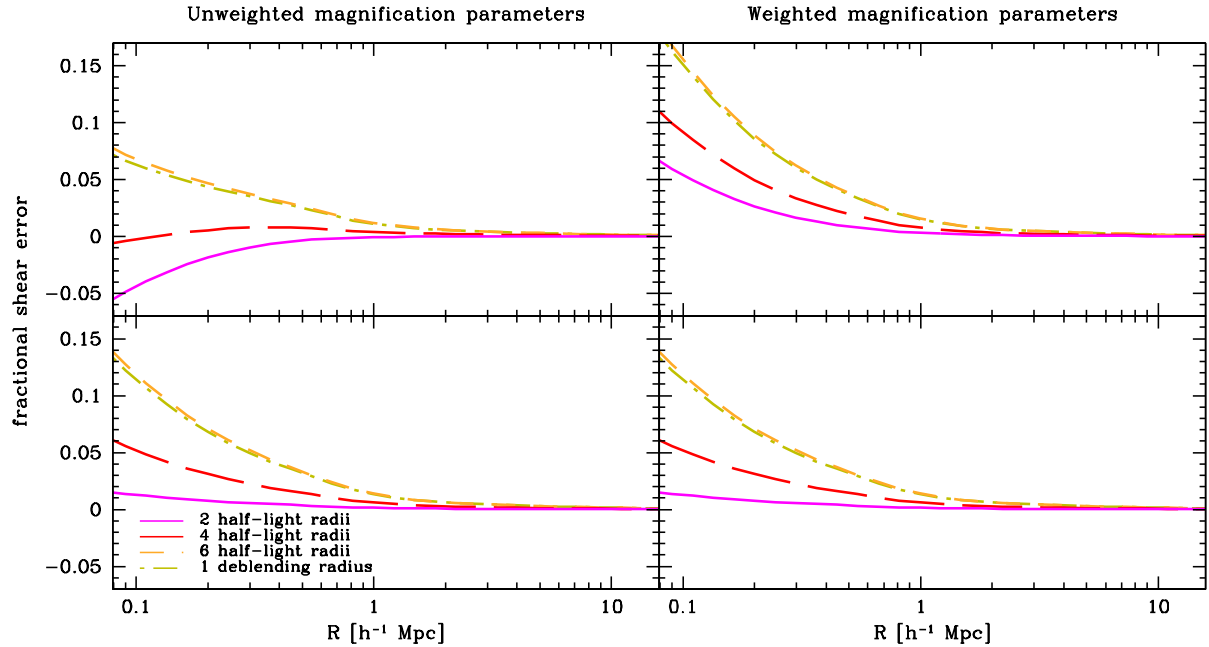} 
  \caption{Fractional errors on the shear in a CFHTLenS-like survey due to neglecting 
            magnification and obscuration (top) or neglecting obscuration only (bottom).  All 
            curves are derived using a scaled version of the SDSS obscuration, 
            Fig.~\ref{fobsc-sdss}, due to the difficulty of measuring the obscuration in CFHTLenS.
            Magnification parameters derived using unweighted galaxy 
            densities are shown in the left-hand column, and using
            weighted densities in the 
            right-hand column.  The magnification is more important in this survey than in SDSS, 
            unsurprising given the depth.  Like SDSS, the obscuration is important in the 
            centers of the clusters.}\label{shear-errors-cfhtlens}
  \end{center}
\end{figure*}

\begin{figure}
  \begin{center}
  \includegraphics[width=0.45\textwidth]{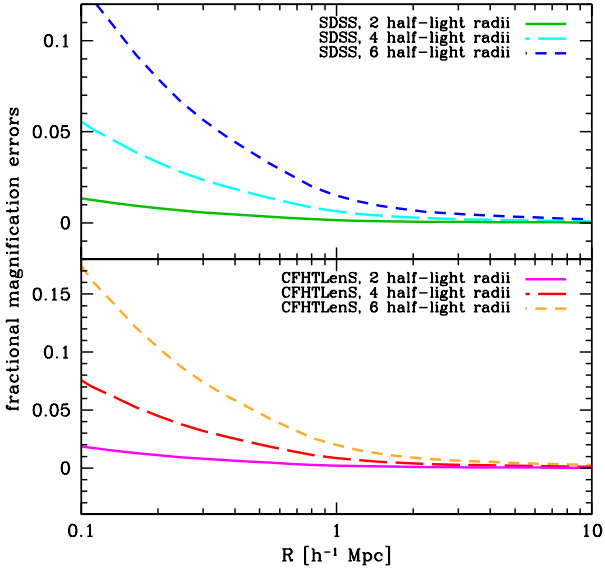} 
  \caption{Fractional errors on the magnification in SDSS-like (top) and CFHTLenS-like (bottom) 
            photometric surveys for a range of obscuration areas per galaxy, assuming that the
            magnification can be measured in narrow annular
            bins.  Since \textit{lens}fit was not used for the photometric catalog, no deblending
            radius is shown for the CFHTLenS sample.}\label{magnification-errors}
  \end{center}
\end{figure}

Since the exact details of the deblender will affect how many half-light radii we can consider the 
galaxy to obscure, we will show results for galaxy areas corresponding to 2, 4 and 6 times the 
half-light radius for SDSS and CFHTLenS, as well as an area corresponding to a radius of 1/2 the 
postage stamp size for CFHTLenS only given the choices made in its deblending process.  Combining the size of an 
average cluster galaxy and the number density of cluster galaxies, we can compute 
$f_{\mathrm{obsc}}(R)$, as shown in Fig.~\ref{fobsc-sdss}.  Assuming
that measurements of stacked lensing are made using number densities
around random points to correct for dilution by physically-assocated
sources, and that magnification estimates likewise neglect the
obscuration effect, we find systematic errors in the shear as 
shown in Figs.~\ref{shear-errors-sdss}~(SDSS) and \ref{shear-errors-cfhtlens}~(CFHTLenS)
and magnification errors as shown in Fig.~\ref{magnification-errors}~(both surveys).

The fraction of the sky obscured by the cluster galaxies depends strongly
on the assumed area  
obscured per galaxy. Even at our most conservative estimate, however, this is a 1 per cent effect for the inner 
regions of GMBCG-like clusters in current surveys.  We would naively expect this problem to increase 
with depth, as 
more galaxies can be detected; however, since most of the galaxies have a size dominated by the 
PSF, a survey with better seeing can counteract this increase by reducing the observed sizes for all 
galaxies, even the ones detected in the shallower survey. By coincidence, these two effects (more 
galaxies but smaller PSF) nearly cancel when moving from SDSS to
CFHTLenS for a given choice of deblend radius, and we see similar
obscuration predictions for both surveys in the bottom right panels of Figures~\ref{shear-errors-sdss}
and \ref{shear-errors-cfhtlens}.  However, since the two surveys do
not make the same choices about deblending, the effect will not
actually be the same for their measurements in practice.

Also of interest is the effect of magnification on the shear errors,
which can be seen by comparing the top and bottom rows in Figures~\ref{shear-errors-sdss}
and \ref{shear-errors-cfhtlens}.  In a CFHTLenS-like survey 
that measures lensing around massive clusters, the magnification can change densities by a few per cent in the inner 
regions of the cluster, which will impact the boost factor and thus the shear by a similar amount.  
Unlike the obscuration, however, this depends on the mass and mass distribution assumed or measured 
for the clusters, so it is not as easy to separate from the measurement of interest.  The difference
between the expected magnification if the parameters are measured using the same shape weights as
the galaxy or not are large as well, highlighting the importance of correctly weighting the number
densities when computing the magnification.

Finally, the obscuration effect can also be seen in a pure magnification measurement.  The effect is larger 
because more galaxies (including cluster galaxies) are detected in the photometric catalogues, where
magnification measurements will be made.  We assume here that the measurement can be
made in annular bins with the same resolution as the shear
measurements.

\begin{figure}
  \begin{center}
  \includegraphics[width=0.45\textwidth]{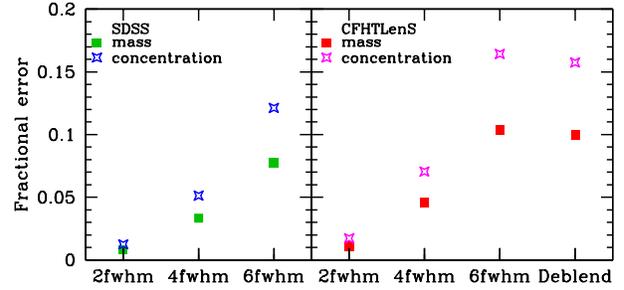} 
  \caption{Fractional errors on the derived mass and concentration for a $10^{14} h^{-1}$~solar 
            mass halo with concentration~$4$, using a two-component NFW fit to false shear data made 
            using a perfect input NFW halo modified by the obscuration measurements of 
            Figs.~\ref{shear-errors-sdss} and \ref{shear-errors-cfhtlens}.  Fits were made in the
            radius range $0.1-2 h^{-1}$~Mpc to avoid the two-halo regime.  One-sigma error bars 
            in logarithmically-spaced annuli were assumed to be
            proportional to $1/r$, as is the case for shape
            noise-dominated errors.  Different choices of radii to use
            or error distributions for weighting the fits 
            will change the impact of this effect.}\label{fit-errors}
  \end{center}
\end{figure}

\begin{figure}
  \begin{center}
  \includegraphics[width=0.45\textwidth]{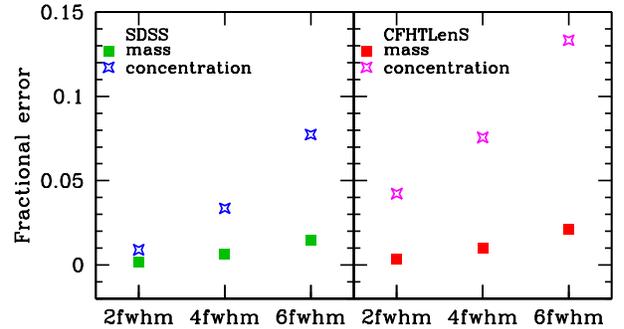} 
  \caption{Fractional errors on the derived mass and concentration for a $10^{14} h^{-1}$~solar 
            mass halo with concentration~$4$, using a two-component NFW fit to false magnification 
            data made using a perfect input NFW halo modified by the obscuration measurements of 
            Fig.~\ref{magnification-errors}.  Fits were made in the
            radius range $0.1-2 h^{-1}$~Mpc to avoid the two-halo regime, and assume that the 
            measurement can be made in narrow annular bins like the shear can.  One-sigma error bars 
            were assumed to be proportional to $1/r$.  Different annulus bins or error distributions
            will change the impact of this effect.}\label{fit-errors-mag}
  \end{center}
\end{figure}

These effects will propagate into measurements of cluster masses and concentrations.  We show the 
mass and concentration errors caused by the (uncorrected) obscuration effect in 
Fig.~\ref{fit-errors} (shear) and Fig.~\ref{fit-errors-mag} (magnification).  These are
fractional errors on both quantities assuming a joint NFW fit to a perfect NFW input.  The effect
is within the error budgets for current measurements with SDSS, but is troublesome for CFHTLenS, 
where there are published cluster masses measured to approximately 10 per cent with 
magnification~\citep{2014MNRAS.439.3755F}.  The mass errors are lessened by approximately 20 per 
cent if we fix the concentration to the input value, since we reduce our sensitivity to the radial 
dependence of the obscuration, but as the concentration may also be a parameter of interest in 
cluster studies we show only the errors from the joint fit.

\section{Forecasts for the impact of obscuration on future surveys}\label{Forecasts}

We can make rough predictions for how future surveys will be impacted by this effect for GMBCG-like 
cluster catalogues. As with our CFHTLenS calculation above, we need only know the average size of a 
PSF-convolved galaxy and the depth of the imaging.  Then we can project the cluster luminosity 
function to various depths in order to predict the galaxy number density and multiply by the area to predict the obscuration 
fraction.  In the case of shear estimates, these predictions are
relevant in the case that the photo-$z$ are sufficiently imperfect
that it is necessary to remove the dilution due to
source galaxies that are physically associated with the lenses (rather
than normalizing by the actual number of background galaxies used).
The estimates of the impact on magnification measurements are more general.

We make predictions for four surveys: the Hyper 
Suprime-Cam survey~\citep[HSC,][]{2012SPIE.8446E..0ZM}; the Dark Energy 
Survey~\citep[DES,][]{2005astro.ph.10346T}, which recently measured this effect for four clusters in
\citet{2014arXiv1405.4285M}; Euclid \citep{2011arXiv1110.3193L}; and the Large 
Synoptic Survey Telescope~\citep[LSST,][]{2009arXiv0912.0201L}.  DES, HSC and LSST are ground-based 
surveys, while Euclid is a space-based survey; DES and HSC are already taking data as of the 
publication date of this paper, while Euclid and LSST will begin in the early 
2020s.  All four of these surveys will engage in large-scale surveys of the sky with 
cosmology as a principal aim, and have both weak lensing and galaxy cluster science among the 
main science drivers of the survey goals.  We summarize the relevant depth and PSF information in 
Table \ref{projection-table}.

\begin{table}
\begin{center}
\begin{tabular}{|l|ll|ll|}
\hline
Survey$\!\!\!$ & Depth$\!\!\!$ & $N_{\mathrm{gals}}/$ & PSF & Projected 
galaxy \\
 & & $N_{\mathrm{gals,SDSS}}\!\!\!$ & FWHM (\arcsec)$\!\!\!$ &  FWHM radius (kpc)$\!\!\!$ \\
\hline
HSC & 25.1 & 3.5 & 0.6 & 2.4 \\
DES & 24.3 & 3 & 0.9 & 3.8 \\
Euclid & 24.5 & 3 & $\leq 0.18$ & 0.73 \\
% 0.14 make note
LSST & 27.5 & 5.5 & 0.7 & 2.8 \\
\hline
\end{tabular}
\end{center}
\caption{Relevant survey parameters for the four surveys we use for projecting the obscuration 
          effect: HSC, DES, Euclid, and LSST.  Depth is reported in the band intended for weak 
          lensing imaging: $i$ for HSC and DES, broad-band visual for Euclid, $r$ or $i$ for LSST.  
          Where available the depth is the depth used in discussions of galaxy detections, generally 
          the 10$\sigma$ extended-source detection limit.  The HSC predictions only report 
          5$\sigma$ detections at a magnitude of 25.9, so we assume a factor of 2 brighter for a 
          10$\sigma$ detection (a more fair comparison to the other surveys) yielding a magnitude 
          of 25.1. 
          The listed Euclid PSF FWHM is the upper limit of the weak
          lensing requirement for the survey as in
          \protect\cite{2011arXiv1110.3193L}, and allows for sources of
          PSF broadening such as jitter and charge diffusion that makes the PSF
          wider than the nominal expectation given the diffraction limit of the telescope and bandpass.
          $N_{\mathrm{gals}}/N_{\mathrm{gals,SDSS}}$ is a prediction based on our fit to the 
          cluster luminosity function, Eq. \ref{clf}, rounded to the nearest half integer. Galaxy 
          radius is a prediction based on the survey PSF, since the
          PSF-convolved size is the relevant quantity
          here.}\label{projection-table}
\end{table}

The average galaxy size will generally be dominated by the PSF, given the power-law distribution of 
galaxy sizes.  However, since we are interested in area rather than radius (which favors larger 
galaxies) and since surveys with smaller PSFs can resolve galaxies below the PSF size for the 
surveys with worse seeing, the average galaxy size is not simply proportional to the FWHM or the 
FWHM squared.  We find that SDSS galaxies have a mean FWHM area approximately $3.5$~times the FWHM area of 
the PSF, and CFHTLenS galaxies have a mean FWHM area approximately $6$~times the FWHM area of the PSF.
For simplicity, we assume upcoming surveys will have galaxy areas $\approx 4$ times the area of
their PSFs.  

\begin{figure}
  \begin{center}
  \includegraphics[width=0.45\textwidth]{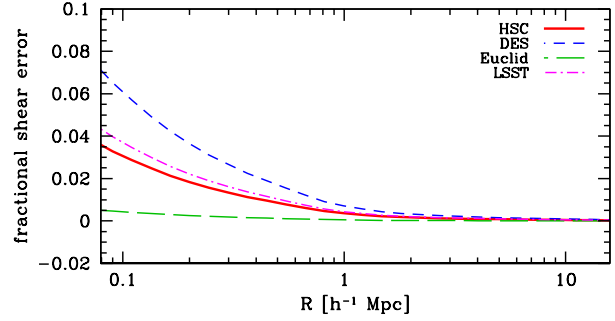} 
  \caption{Estimates of the fractional systematic errors on the shear
    in several upcoming surveys due to neglect of the
            obscuration effect around GMBCG-like clusters. The number of galaxies in the cluster is
            the number in the SDSS catalogue scaled according to the cluster luminosity function,
            Eq.~\ref{clf}; the deblending-relevant average galaxy radius is
            assumed to be 8 times the survey PSF FWHM, which is a typical number based on the 
            results from SDSS and CFHTLenS. The surveys are close enough in depth that the PSF
            size dominates, giving DES the largest errors due to this effect and Euclid, a 
            space-based survey, the smallest.}\label{shear-errors-future}
  \end{center}
\end{figure}

We can rescale the obscuration signal around GMBCG clusters in SDSS by
these numbers.  Fig. \ref{shear-errors-future} shows the systematic
errors on the shear for
future surveys that measure cluster lensing without accounting for
obscuration, assuming a galaxy deblend area with a radius $4$~times
the galaxy FWHM radius 
($8$~times the PSF FWHM) and a GMBCG-like mass and redshift distribution.  The effect is worst in 
DES, given the combination of a deep survey with a larger PSF. It is hardly detectable in Euclid 
given the small space-based PSF.  The shape of the PSF and not its FWHM may be more important for
diffraction-limited space-based surveys, which could alter the predicted size of the effect, but our
prediction for the size of this effect is so small that even factors of
$\sim 2$ changes would not alter the essential conclusion that this
effect is not a substantial source of systematic error for Euclid
lensing analysis of clusters in the $\sim 10^{14}h^{-1}M_\odot$ mass range.

All the ground-based surveys have a maximum shear error
of $\gtrsim$~5 per cent at the minimum radius we consider ($0.1h^{-1}$Mpc).  We can 
see from Fig.~\ref{fit-errors} that the error on the mass is of comparable size to the maximum 
shear error, so neglecting the obscuration effect may impact cluster mass estimates to nearly that 
level--an important effect for all of these surveys given that they are trying to push stacked cluster mass measurements 
below 10 per cent errors.

We can compare this prediction to the effect that has recently been observed in the Dark Energy
Survey by \citet{2014arXiv1405.4285M}.  The galaxy clusters studied in that work are slightly 
lower-redshift than the GMBCG clusters (redshifts $\approx 0.3-0.4$ for three of the clusters,
versus the peak GMBCG redshift of ~0.45; the fourth cluster is the nearby Abell 3216).  They
are also more massive\footnote{While no mass measurements have been
made for GMBCG, the similar MaxBCG catalog \citep{2007ApJ...660..239K} contains clusters
with a typical mass of $\approx 10^{14} h^{-1}$ solar masses
\citep{2008JCAP...08..006M}.  Both catalogs have a similar comoving number 
density of approximately $2\times 10^{-5} h^3 \mathrm{Mpc}^-3$, and transforming the richness
estimator of GMBCG into the MaxBCG richness estimator and applying the scaling relation of
\citet{2008JCAP...08..006M} also yields cluster masses of this order.}, 
with the three distant clusters all $\geq 10^{15} h^{-1}$ solar masses. 
 The nearness of the clusters will
slightly counteract the obscuration effect (since galaxies convolved with the same angular size
PSF will look smaller in physical units at smaller $z$).  The mass will increase it, though, by
adding more cluster galaxies.  The mass is usually found to be
proportional to the number of galaxies above a luminosity threshhold
as approximately $N_{\mathrm{gals}}^{1.1-1.2}$, so we can expect in
turn that the number of galaxies will be proportional to
$M_{\mathrm{cl}}^{0.8-0.9}$, giving us a factor of $\sim 6-8$ in
number of galaxies and thus obscuration, or 50 to 70 per cent
obscuration at $0.1 h^{-1}$ Mpc ignoring redshift effects, while
\citeauthor{2014arXiv1405.4285M} report a criterion equivalent to
approximately 40 to 60 per cent obscuration as well, a reasonable level of agreement given the
degree of estimation involved.

The effect is also measured differently in \citeauthor{2014arXiv1405.4285M}.  They measure the
obscured fraction of the sky directly, by using the BALROG 
software\footnote{\url{https://github.com/emhuff/Balrog}} to add
a simulated galaxy to the measured data and then checking if the pipeline detects the new
galaxy or not.  This also captures other effects such as problems in background subtraction 
in dense regions, and is probably the most formally correct way to solve the problem.  However,
once the average deblending size of galaxies as a function of redshift is measured, our method 
of predicting the impact of obscuration on cluster shear and
magnification measurements essentially comes for free with any
$\Delta\Sigma$ computation, and may be more computationally 
feasible for large, heterogeneous cluster samples.

\section{Conclusions}

We have presented an analysis of the impact on weak lensing shear and magnification measurements
of galaxy clusters due to the obscuration of the background sky caused by galaxies within those same
clusters due to the difficulty of separating nearby galaxies.  The effect is small enough to be 
within the error budgets
of current surveys, with the possible exception of CFHTLenS given its conservative deblending choices, but it is
large enough to become important as deeper surveys try to reach better precision on their galaxy
cluster mass estimates.  Thus, schemes for accounting for this effect,
such as that used by \cite{2014arXiv1405.4285M}, should be
incorporated into the cluster lensing measurements of future surveys.
Since the effect is a function of radius, it impacts both the mass and  
concentration estimates for the clusters in nonlinear ways.

The precise size of the effect is difficult to compute.  For our method, the number of cluster 
galaxies must be 
measured, which is not difficult as long as the obscuration leaves a majority of the objects
measured.  However, the average area covered by a galaxy is subtle, as it relies on the details of
the galaxy light distribution, the effect of the PSF convolution, and the success of the deblending
software at disentangling nearby objects.  Detailed anaylsis of deblender behavior on realistic 
galaxies will be required to compute the size of this effect in real data.  However, a rough 
estimate of the effect size, based on simple arguments, shows that obscuration will be important for
deeper surveys with more precise mass measurement targets.  Corrections for obscuration will be 
necessary for upcoming surveys to meet their requirements on cluster
mass estimates using stacked lensing and clustering measurements.

\section{Acknowledgements}

This work was supported by the Department of Energy Early Career Award program.

We thank Ying Zu, Mike Jarvis, Eduardo Rozo, Fabian Schmidt, Ross O'Connell, and Mariana 
Vargas-Marga\~{n}a for useful discussions related to this work.

Funding for the SDSS and SDSS-II has been provided by the Alfred P. Sloan Foundation, the 
Participating Institutions, the National Science Foundation, the U.S. Department of Energy, the 
National Aeronautics and Space Administration, the Japanese Monbukagakusho, the Max Planck Society, 
and the Higher Education Funding Council for England. The SDSS Web Site is http://www.sdss.org/.  
The SDSS is managed by the Astrophysical Research Consortium for the Participating Institutions. 
The Participating Institutions are the American Museum of Natural History, Astrophysical Institute 
Potsdam, University of Basel, University of Cambridge, Case Western Reserve University, University 
of Chicago, Drexel University, Fermilab, the Institute for Advanced Study, the Japan Participation 
Group, Johns Hopkins University, the Joint Institute for Nuclear Astrophysics, the Kavli Institute 
for Particle Astrophysics and Cosmology, the Korean Scientist Group, the Chinese Academy of Sciences 
(LAMOST), Los Alamos National Laboratory, the Max-Planck-Institute for Astronomy (MPIA), the 
Max-Planck-Institute for Astrophysics (MPA), New Mexico State University, Ohio State University, 
University of Pittsburgh, University of Portsmouth, Princeton University, the United States Naval 
Observatory, and the University of Washington.

This work is based on observations obtained with MegaPrime/MegaCam, a joint project of CFHT and 
CEA/IRFU, at the Canada-France-Hawaii Telescope (CFHT) which is operated by the National Research 
Council (NRC) of Canada, the Institut National des Sciences de l'Univers of the Centre National de 
la Recherche Scientifique (CNRS) of France, and the University of Hawaii. This research used the 
facilities of the Canadian Astronomy Data Centre operated by the National Research Council of Canada 
with the support of the Canadian Space Agency. CFHTLenS data processing was made possible thanks to 
significant computing support from the NSERC Research Tools and Instruments grant program.

\bibliographystyle{mn2e}
\bibliography{obscuration}

\begin{thebibliography}{}

\bibitem[\protect\citeauthoryear{{Aihara} et~al.,}{{Aihara}
  et~al.}{2011}]{2011ApJS..193...29A}
{Aihara} H.,  et~al., 2011, ApJs, 193, 29

\bibitem[\protect\citeauthoryear{{Albrecht}, {Bernstein}, {Cahn}, {Freedman},
  {Hewitt}, {Hu}, {Huth}, {Kamionkowski}, {Kolb}, {Knox}, {Mather}, {Staggs} \&
  {Suntzeff}}{{Albrecht} et~al.}{2006}]{2006astro.ph..9591A}
{Albrecht} A.,  {Bernstein} G.,  {Cahn} R.,  {Freedman} W.~L.,  {Hewitt} J.,
  {Hu} W.,  {Huth} J.,  {Kamionkowski} M.,  {Kolb} E.~W.,  {Knox} L.,  {Mather}
  J.~C.,  {Staggs} S.,    {Suntzeff} N.~B.,  2006, ArXiv Astrophysics e-prints

\bibitem[\protect\citeauthoryear{{Allen}, {Evrard} \& {Mantz}}{{Allen}
  et~al.}{2011}]{2011ARAA..49..409A}
{Allen} S.~W.,  {Evrard} A.~E.,    {Mantz} A.~B.,  2011, ARA\&A, 49, 409

\bibitem[\protect\citeauthoryear{{Applegate}, {von der Linden}, {Kelly},
  {Allen}, {Allen}, {Burchat}, {Burke}, {Ebeling}, {Mantz} \&
  {Morris}}{{Applegate} et~al.}{2014}]{2014MNRAS.439...48A}
{Applegate} D.~E.,  {von der Linden} A.,  {Kelly} P.~L.,  {Allen} M.~T.,
  {Allen} S.~W.,  {Burchat} P.~R.,  {Burke} D.~L.,  {Ebeling} H.,  {Mantz} A.,
    {Morris} R.~G.,  2014, MNRAS, 439, 48

\bibitem[\protect\citeauthoryear{{Bartelmann} \& {Schneider}}{{Bartelmann} \&
  {Schneider}}{2001}]{2001PhR...340..291B}
{Bartelmann} M.,  {Schneider} P.,  2001, Phys. Rep., 340, 291

\bibitem[\protect\citeauthoryear{{Ben{\'{\i}}tez}}{{Ben{\'{\i}}tez}}{2000}]{20%
00ApJ...536..571B}
{Ben{\'{\i}}tez} N.,  2000, ApJ, 536, 571

\bibitem[\protect\citeauthoryear{{Bernstein} \& {Huterer}}{{Bernstein} \&
  {Huterer}}{2010}]{2010MNRAS.401.1399B}
{Bernstein} G.,  {Huterer} D.,  2010, MNRAS, 401, 1399

\bibitem[\protect\citeauthoryear{{Bernstein} \& {Jarvis}}{{Bernstein} \&
  {Jarvis}}{2002}]{2002AJ....123..583B}
{Bernstein} G.~M.,  {Jarvis} M.,  2002, AJ, 123, 583

\bibitem[\protect\citeauthoryear{{Bertin} \& {Arnouts}}{{Bertin} \&
  {Arnouts}}{1996}]{1996AAS..117..393B}
{Bertin} E.,  {Arnouts} S.,  1996, A\&AS, 117, 393

\bibitem[\protect\citeauthoryear{{Blazek}, {Mandelbaum}, {Seljak} \&
  {Nakajima}}{{Blazek} et~al.}{2012}]{2012JCAP...05..041B}
{Blazek} J.,  {Mandelbaum} R.,  {Seljak} U.,    {Nakajima} R.,  2012, JCAP, 5,
  41

\bibitem[\protect\citeauthoryear{{Bridle} et~al.,}{{Bridle}
  et~al.}{2010}]{2010MNRAS.405.2044B}
{Bridle} S.,  et~al., 2010, MNRAS, 405, 2044

\bibitem[\protect\citeauthoryear{{Capak}}{{Capak}}{2004}]{2004PhDT.........4C}
{Capak} P.~L.,  2004, PhD thesis, UNIVERSITY OF HAWAI'I

\bibitem[\protect\citeauthoryear{{Coe}, {Ben{\'{\i}}tez}, {S{\'a}nchez}, {Jee},
  {Bouwens} \& {Ford}}{{Coe} et~al.}{2006}]{2006AJ....132..926C}
{Coe} D.,  {Ben{\'{\i}}tez} N.,  {S{\'a}nchez} S.~F.,  {Jee} M.,  {Bouwens} R.,
     {Ford} H.,  2006, AJ, 132, 926

\bibitem[\protect\citeauthoryear{{Corless} \& {King}}{{Corless} \&
  {King}}{2007}]{2007MNRAS.380..149C}
{Corless} V.~L.,  {King} L.~J.,  2007, MNRAS, 380, 149

\bibitem[\protect\citeauthoryear{{Erben} et~al.,}{{Erben}
  et~al.}{2013}]{2013MNRAS.433.2545E}
{Erben} T.,  et~al., 2013, MNRAS, 433, 2545

\bibitem[\protect\citeauthoryear{{Feldmann} et~al.,}{{Feldmann}
  et~al.}{2006}]{2006MNRAS.372..565F}
{Feldmann} R.,  et~al., 2006, MNRAS, 372, 565

\bibitem[\protect\citeauthoryear{{Ford}, {Hildebrandt}, {Van Waerbeke},
  {Erben}, {Laigle}, {Milkeraitis} \& {Morrison}}{{Ford}
  et~al.}{2014}]{2014MNRAS.439.3755F}
{Ford} J.,  {Hildebrandt} H.,  {Van Waerbeke} L.,  {Erben} T.,  {Laigle} C.,
  {Milkeraitis} M.,    {Morrison} C.~B.,  2014, MNRAS, 439, 3755

\bibitem[\protect\citeauthoryear{{George}, {Leauthaud}, {Bundy}, {Finoguenov},
  {Ma}, {Rykoff}, {Tinker}, {Wechsler}, {Massey} \& {Mei}}{{George}
  et~al.}{2012}]{2012ApJ...757....2G}
{George} M.~R.,  {Leauthaud} A.,  {Bundy} K.,  {Finoguenov} A.,  {Ma} C.-P.,
  {Rykoff} E.~S.,  {Tinker} J.~L.,  {Wechsler} R.~H.,  {Massey} R.,    {Mei}
  S.,  2012, ApJ, 757, 2

\bibitem[\protect\citeauthoryear{{Hamana}, {Oguri}, {Shirasaki} \&
  {Sato}}{{Hamana} et~al.}{2012}]{2012MNRAS.425.2287H}
{Hamana} T.,  {Oguri} M.,  {Shirasaki} M.,    {Sato} M.,  2012, MNRAS, 425,
  2287

\bibitem[\protect\citeauthoryear{{Hao}, {McKay}, {Koester}, {Rykoff}, {Rozo},
  {Annis}, {Wechsler}, {Evrard}, {Siegel}, {Becker}, {Busha}, {Gerdes},
  {Johnston} \& {Sheldon}}{{Hao} et~al.}{2010}]{2010ApJS..191..254H}
{Hao} J.,  {McKay} T.~A.,  {Koester} B.~P.,  {Rykoff} E.~S.,  {Rozo} E.,
  {Annis} J.,  {Wechsler} R.~H.,  {Evrard} A.,  {Siegel} S.~R.,  {Becker} M.,
  {Busha} M.,  {Gerdes} D.,  {Johnston} D.~E.,    {Sheldon} E.,  2010, ApJs,
  191, 254

\bibitem[\protect\citeauthoryear{{Heymans} et~al.,}{{Heymans}
  et~al.}{2012}]{2012MNRAS.427..146H}
{Heymans} C.,  et~al., 2012, MNRAS, 427, 146

\bibitem[\protect\citeauthoryear{{Heymans} et~al.,}{{Heymans}
  et~al.}{2013}]{2013MNRAS.432.2433H}
{Heymans} C.,  et~al., 2013, MNRAS, 432, 2433

\bibitem[\protect\citeauthoryear{{Hildebrandt} et~al.,}{{Hildebrandt}
  et~al.}{2012}]{2012MNRAS.421.2355H}
{Hildebrandt} H.,  et~al., 2012, MNRAS, 421, 2355

\bibitem[\protect\citeauthoryear{{Hirata} \& {Seljak}}{{Hirata} \&
  {Seljak}}{2003}]{2003MNRAS.343..459H}
{Hirata} C.,  {Seljak} U.,  2003, MNRAS, 343, 459

\bibitem[\protect\citeauthoryear{{Hirata} \& {Seljak}}{{Hirata} \&
  {Seljak}}{2004}]{2004PhRvD..70f3526H}
{Hirata} C.~M.,  {Seljak} U.,  2004, PRD, 70, 063526

\bibitem[\protect\citeauthoryear{{Hoekstra}, {Hartlap}, {Hilbert} \& {van
  Uitert}}{{Hoekstra} et~al.}{2011}]{2011MNRAS.412.2095H}
{Hoekstra} H.,  {Hartlap} J.,  {Hilbert} S.,    {van Uitert} E.,  2011, MNRAS,
  412, 2095

\bibitem[\protect\citeauthoryear{{Kitching} et~al.,}{{Kitching}
  et~al.}{2012}]{2012MNRAS.423.3163K}
{Kitching} T.~D.,  et~al., 2012, MNRAS, 423, 3163

\bibitem[\protect\citeauthoryear{{Koester} et~al.,}{{Koester}
  et~al.}{2007}]{2007ApJ...660..239K}
{Koester} B.~P.,  et~al., 2007, ApJ, 660, 239

\bibitem[\protect\citeauthoryear{{Kravtsov} \& {Borgani}}{{Kravtsov} \&
  {Borgani}}{2012}]{2012ARAA..50..353K}
{Kravtsov} A.~V.,  {Borgani} S.,  2012, ARA\&A, 50, 353

\bibitem[\protect\citeauthoryear{{Laureijs}, {Amiaux}, {Arduini},
  {Augu{\`e}res}, {Brinchmann}, {Cole}, {Cropper}, {Dabin}, {Duvet}, {Ealet} \&
  et al.}{{Laureijs} et~al.}{2011}]{2011arXiv1110.3193L}
{Laureijs} R.,  {Amiaux} J.,  {Arduini} S.,  {Augu{\`e}res} J.~.,  {Brinchmann}
  J.,  {Cole} R.,  {Cropper} M.,  {Dabin} C.,  {Duvet} L.,  {Ealet} A.,    et
  al. 2011, ArXiv e-prints

\bibitem[\protect\citeauthoryear{{LSST Science Collaboration}, {Abell},
  {Allison}, {Anderson}, {Andrew}, {Angel}, {Armus}, {Arnett}, {Asztalos},
  {Axelrod} \& et al.}{{LSST Science Collaboration}
  et~al.}{2009}]{2009arXiv0912.0201L}
{LSST Science Collaboration} {Abell} P.~A.,  {Allison} J.,  {Anderson} S.~F.,
  {Andrew} J.~R.,  {Angel} J.~R.~P.,  {Armus} L.,  {Arnett} D.,  {Asztalos}
  S.~J.,  {Axelrod} T.~S.,    et al. 2009, ArXiv e-prints

\bibitem[\protect\citeauthoryear{{Lupton}, {Gunn}, {Ivezi{\'c}}, {Knapp} \&
  {Kent}}{{Lupton} et~al.}{2001}]{2001ASPC..238..269L}
{Lupton} R.,  {Gunn} J.~E.,  {Ivezi{\'c}} Z.,  {Knapp} G.~R.,    {Kent} S.,
  2001, in {Harnden} Jr. F.~R.,  {Primini} F.~A.,   {Payne} H.~E.,  eds,
  Astronomical Data Analysis Software and Systems X Vol.~238 of Astronomical
  Society of the Pacific Conference Series, {The SDSS Imaging Pipelines}.
p.~269

\bibitem[\protect\citeauthoryear{{Mandelbaum}, {Hirata}, {Seljak}, {Guzik},
  {Padmanabhan}, {Blake}, {Blanton}, {Lupton} \& {Brinkmann}}{{Mandelbaum}
  et~al.}{2005}]{2005MNRAS.361.1287M}
{Mandelbaum} R.,  {Hirata} C.~M.,  {Seljak} U.,  {Guzik} J.,  {Padmanabhan} N.,
   {Blake} C.,  {Blanton} M.~R.,  {Lupton} R.,    {Brinkmann} J.,  2005, MNRAS,
  361, 1287

\bibitem[\protect\citeauthoryear{{Mandelbaum}, {Seljak}, {Baldauf} \&
  {Smith}}{{Mandelbaum} et~al.}{2010}]{2010MNRAS.405.2078M}
{Mandelbaum} R.,  {Seljak} U.,  {Baldauf} T.,    {Smith} R.~E.,  2010, MNRAS,
  405, 2078

\bibitem[\protect\citeauthoryear{{Mandelbaum}, {Seljak}, {Cool}, {Blanton},
  {Hirata} \& {Brinkmann}}{{Mandelbaum} et~al.}{2006}]{2006MNRAS.372..758M}
{Mandelbaum} R.,  {Seljak} U.,  {Cool} R.~J.,  {Blanton} M.,  {Hirata} C.~M.,
   {Brinkmann} J.,  2006, MNRAS, 372, 758

\bibitem[\protect\citeauthoryear{{Mandelbaum}, {Seljak} \&
  {Hirata}}{{Mandelbaum} et~al.}{2008}]{2008JCAP...08..006M}
{Mandelbaum} R.,  {Seljak} U.,    {Hirata} C.~M.,  2008, JCAP, 8, 6

\bibitem[\protect\citeauthoryear{{Marian}, {Smith} \& {Bernstein}}{{Marian}
  et~al.}{2010}]{2010ApJ...709..286M}
{Marian} L.,  {Smith} R.~E.,    {Bernstein} G.~M.,  2010, ApJ, 709, 286

\bibitem[\protect\citeauthoryear{{Massey}, {Hoekstra}, {Kitching}, {Rhodes},
  {Cropper}, {Amiaux}, {Harvey}, {Mellier}, {Meneghetti}, {Miller},
  {Paulin-Henriksson}, {Pires}, {Scaramella} \& {Schrabback}}{{Massey}
  et~al.}{2013}]{2013MNRAS.429..661M}
{Massey} R.,  {Hoekstra} H.,  {Kitching} T.,  {Rhodes} J.,  {Cropper} M.,
  {Amiaux} J.,  {Harvey} D.,  {Mellier} Y.,  {Meneghetti} M.,  {Miller} L.,
  {Paulin-Henriksson} S.,  {Pires} S.,  {Scaramella} R.,    {Schrabback} T.,
  2013, MNRAS, 429, 661

\bibitem[\protect\citeauthoryear{{Melchior} et~al.,}{{Melchior}
  et~al.}{2014}]{2014arXiv1405.4285M}
{Melchior} P.,  et~al., 2014, ArXiv e-prints

\bibitem[\protect\citeauthoryear{{Miller} et~al.,}{{Miller}
  et~al.}{2013}]{2013MNRAS.429.2858M}
{Miller} L.,  et~al., 2013, MNRAS, 429, 2858

\bibitem[\protect\citeauthoryear{{Miyazaki} et~al.,}{{Miyazaki}
  et~al.}{2012}]{2012SPIE.8446E..0ZM}
{Miyazaki} S.,  et~al., 2012, in Society of Photo-Optical Instrumentation
  Engineers (SPIE) Conference Series Vol.~8446 of Society of Photo-Optical
  Instrumentation Engineers (SPIE) Conference Series, {Hyper Suprime-Cam}

\bibitem[\protect\citeauthoryear{{Nakajima}, {Mandelbaum}, {Seljak}, {Cohn},
  {Reyes} \& {Cool}}{{Nakajima} et~al.}{2012}]{2012MNRAS.420.3240N}
{Nakajima} R.,  {Mandelbaum} R.,  {Seljak} U.,  {Cohn} J.~D.,  {Reyes} R.,
  {Cool} R.,  2012, MNRAS, 420, 3240

\bibitem[\protect\citeauthoryear{{Parker}, {Hoekstra}, {Hudson}, {van Waerbeke}
  \& {Mellier}}{{Parker} et~al.}{2007}]{2007ApJ...669...21P}
{Parker} L.~C.,  {Hoekstra} H.,  {Hudson} M.~J.,  {van Waerbeke} L.,
  {Mellier} Y.,  2007, ApJ, 669, 21

\bibitem[\protect\citeauthoryear{{Pracy}, {Driver}, {De Propris}, {Couch} \&
  {Nulsen}}{{Pracy} et~al.}{2005}]{2005MNRAS.364.1147P}
{Pracy} M.~B.,  {Driver} S.~P.,  {De Propris} R.,  {Couch} W.~J.,    {Nulsen}
  P.~E.~J.,  2005, MNRAS, 364, 1147

\bibitem[\protect\citeauthoryear{{Reyes}, {Mandelbaum}, {Gunn}, {Nakajima},
  {Seljak} \& {Hirata}}{{Reyes} et~al.}{2012}]{2012MNRAS.425.2610R}
{Reyes} R.,  {Mandelbaum} R.,  {Gunn} J.~E.,  {Nakajima} R.,  {Seljak} U.,
  {Hirata} C.~M.,  2012, MNRAS, 425, 2610

\bibitem[\protect\citeauthoryear{{Rozo}, {Wu} \& {Schmidt}}{{Rozo}
  et~al.}{2011}]{2011ApJ...735..118R}
{Rozo} E.,  {Wu} H.-Y.,    {Schmidt} F.,  2011, ApJ, 735, 118

\bibitem[\protect\citeauthoryear{{Schechter}}{{Schechter}}{1976}]{1976ApJ...20%
3..297S}
{Schechter} P.,  1976, ApJ, 203, 297

\bibitem[\protect\citeauthoryear{{Schmidt}, {Rozo}, {Dodelson}, {Hui} \&
  {Sheldon}}{{Schmidt} et~al.}{2009}]{2009PhRvL.103e1301S}
{Schmidt} F.,  {Rozo} E.,  {Dodelson} S.,  {Hui} L.,    {Sheldon} E.,  2009,
  Physical Review Letters, 103, 051301

\bibitem[\protect\citeauthoryear{{Schneider}, {Frenk} \& {Cole}}{{Schneider}
  et~al.}{2012}]{2012JCAP...05..030S}
{Schneider} M.~D.,  {Frenk} C.~S.,    {Cole} S.,  2012, JCAP, 5, 30

\bibitem[\protect\citeauthoryear{{Sheldon}, {Cunha}, {Mandelbaum}, {Brinkmann}
  \& {Weaver}}{{Sheldon} et~al.}{2012}]{2012ApJS..201...32S}
{Sheldon} E.~S.,  {Cunha} C.~E.,  {Mandelbaum} R.,  {Brinkmann} J.,    {Weaver}
  B.~A.,  2012, ApJS, 201, 32

\bibitem[\protect\citeauthoryear{{Sheldon}, {Johnston}, {Frieman}, {Scranton},
  {McKay}, {Connolly}, {Budav{\'a}ri}, {Zehavi}, {Bahcall}, {Brinkmann} \&
  {Fukugita}}{{Sheldon} et~al.}{2004}]{2004AJ....127.2544S}
{Sheldon} E.~S.,  {Johnston} D.~E.,  {Frieman} J.~A.,  {Scranton} R.,  {McKay}
  T.~A.,  {Connolly} A.~J.,  {Budav{\'a}ri} T.,  {Zehavi} I.,  {Bahcall} N.~A.,
   {Brinkmann} J.,    {Fukugita} M.,  2004, AJ, 127, 2544

\bibitem[\protect\citeauthoryear{{Sheldon}, {Johnston}, {Scranton}, {Koester},
  {McKay}, {Oyaizu}, {Cunha}, {Lima}, {Lin}, {Frieman}, {Wechsler}, {Annis},
  {Mandelbaum}, {Bahcall} \& {Fukugita}}{{Sheldon}
  et~al.}{2009}]{2009ApJ...703.2217S}
{Sheldon} E.~S.,  {Johnston} D.~E.,  {Scranton} R.,  {Koester} B.~P.,  {McKay}
  T.~A.,  {Oyaizu} H.,  {Cunha} C.,  {Lima} M.,  {Lin} H.,  {Frieman} J.~A.,
  {Wechsler} R.~H.,  {Annis} J.,  {Mandelbaum} R.,  {Bahcall} N.~A.,
  {Fukugita} M.,  2009, ApJ, 703, 2217

\bibitem[\protect\citeauthoryear{{Stoughton} et~al.,}{{Stoughton}
  et~al.}{2002}]{2002AJ....123..485S}
{Stoughton} C.,  et~al., 2002, AJ, 123, 485

\bibitem[\protect\citeauthoryear{{Swanson}, {Tegmark}, {Hamilton} \&
  {Hill}}{{Swanson} et~al.}{2008}]{2008MNRAS.387.1391S}
{Swanson} M.~E.~C.,  {Tegmark} M.,  {Hamilton} A.~J.~S.,    {Hill} J.~C.,
  2008, MNRAS, 387, 1391

\bibitem[\protect\citeauthoryear{{The Dark Energy Survey Collaboration}}{{The
  Dark Energy Survey Collaboration}}{2005}]{2005astro.ph.10346T}
{The Dark Energy Survey Collaboration} 2005, ArXiv Astrophysics e-prints

\bibitem[\protect\citeauthoryear{{Voit}}{{Voit}}{2005}]{2005RvMP...77..207V}
{Voit} G.~M.,  2005, Reviews of Modern Physics, 77, 207

\bibitem[\protect\citeauthoryear{{Weinberg}, {Mortonson}, {Eisenstein},
  {Hirata}, {Riess} \& {Rozo}}{{Weinberg} et~al.}{2013}]{2013PhR...530...87W}
{Weinberg} D.~H.,  {Mortonson} M.~J.,  {Eisenstein} D.~J.,  {Hirata} C.,
  {Riess} A.~G.,    {Rozo} E.,  2013, Phys. Rep., 530, 87

\bibitem[\protect\citeauthoryear{{Wright} \& {Brainerd}}{{Wright} \&
  {Brainerd}}{2000}]{2000ApJ...534...34W}
{Wright} C.~O.,  {Brainerd} T.~G.,  2000, ApJ, 534, 34

\bibitem[\protect\citeauthoryear{{York} et~al.,}{{York}
  et~al.}{2000}]{2000AJ....120.1579Y}
{York} D.~G.,  et~al., 2000, AJ, 120, 1579

\bibitem[\protect\citeauthoryear{{Zenteno} et~al.,}{{Zenteno}
  et~al.}{2011}]{2011ApJ...734....3Z}
{Zenteno} A.,  et~al., 2011, ApJ, 734, 3

\end{thebibliography}

\end{document}